\documentclass[a4paper,fleqn]{cas-dc}

\usepackage[numbers]{natbib}
\usepackage{comment}
\usepackage{float}
\usepackage{bm}
\usepackage{makecell}
\usepackage{notoccite}
\usepackage{siunitx}
\usepackage{url}
\usepackage{pgfplots}
\pgfplotsset{compat=1.5}
\usepackage{tikz}
\usetikzlibrary{pgfplots.dateplot,}
\usetikzlibrary{matrix,calc}
\usetikzlibrary{calc}
\usetikzlibrary{positioning}
\usetikzlibrary{shapes.geometric, arrows}

\tikzstyle{Detector} = [rectangle, rounded corners, minimum width=2cm, minimum height=1.5cm,text centered, draw=black, fill=cornflowerblue!30]

\tikzstyle{Sampler} = [rectangle, rounded corners, minimum width=2cm, minimum height=1.5cm,text centered, draw=black, fill=lightseagreen!30]

\tikzstyle{Classifier} = [rectangle, rounded corners, minimum width=2cm, minimum height=1.5cm,text centered, draw=black, fill=sandybrown!30]

\tikzstyle{Frame} = [rectangle, minimum width=2cm, minimum height=1.5cm,text centered, draw=black, fill=none]

\tikzstyle{arrow} = [thick,->,>=stealth]

\tikzstyle{arrow_small} = [thick,densely dotted,->,>=stealth]

\usepackage{algorithm}
\usepackage[noend]{algpseudocode}

\definecolor{color1}{RGB}{0,0,0} 
\definecolor{color2}{RGB}{182,193,251} 
\definecolor{bblue}{HTML}{4F81BD}
\definecolor{rred}{HTML}{C0504D}
\definecolor{ggreen}{HTML}{9BBB59}
\definecolor{ppurple}{HTML}{9F4C7C}
\definecolor{sandybrown}{HTML}{f4a460}
\definecolor{lightseagreen}{HTML}{20b2aa}
\definecolor{cornflowerblue}{HTML}{6495ed}
\definecolor{limegreen}{HTML}{32CD32}
\definecolor{orange}{HTML}{ffa500}
\definecolor{purple}{HTML}{6495ed}

\usepackage[automake,acronym,nomain,nopostdot=true,nonumberlist]{glossaries-extra}
\usepackage{glossary-mcols}
\setabbreviationstyle[acronym]{long-short}
\GlsXtrEnableEntryCounting{acronym}{1}
\newacronym{DSO}{DSO}{distribution system operator}
\newacronym{TSO}{TSO}{transmission system operator}
\newacronym{FA}{FA}{flexibility activation}
\newacronym{EIP}{EIP}{event identification pipeline}
\newacronym{DG}{DN}{distribution network}
\newacronym{ICT}{ICT}{information and communication technology}
\newacronym{SM}{SM}{smart meter}
\newacronym{mPMU}{$\mu$PMU}{micro phasor measurement unit}
\newacronym{OS}{OS}{open-set}
\newacronym{CS}{CS}{closed-set}
\newacronym{HTM}{HTM}{hierarchical temporal memory}
\newacronym{ARIMA}{ARIMA}{autoregressive integrated moving average}
\newacronym{CNN}{CNN}{convolutional neural network}
\newacronym{SR}{SR}{spectral residual}
\newacronym{EVM}{EVM}{extreme value machine}
\newacronym{TP}{TP}{true positive}
\newacronym{FP}{FP}{false positive}
\newacronym{FN}{FN}{false negative}
\newacronym{TN}{TN}{true negative}
\newacronym{AIC}{AIC}{Akaike information criterion}
\newacronym{NO}{NO}{normal operation}
\newacronym{MP}{MP}{Monday peak}
\newacronym{DU}{DU}{data unavailability}
\newacronym{FV}{FV}{frozen value}
\newacronym{DER}{DER}{distributed energy resource}

\makeglossaries

\def\tsc#1{\csdef{#1}{\textsc{\lowercase{#1}}\xspace}}
\tsc{WGM}
\tsc{QE}
\tsc{EP}
\tsc{PMS}
\tsc{BEC}
\tsc{DE}
\begin{document}

\let\WriteBookmarks\relax
\def\floatpagepagefraction{1}
\def\textpagefraction{.001}
\shorttitle{}
\shortauthors{Müller et~al.}

\title [mode = title]{Unsupervised detection and open-set classification of fast-ramped flexibility activation events}                      

\author[1]{Nils Müller}[type=editor,
                        auid=000,bioid=1,
                        orcid=0000-0002-3749-5073
                        ]
\cormark[1]

\ead{nilmu@elektro.dtu.dk}

\address[1]{Center for Electric Power and Energy, Technical University of Denmark, Elektrovej, Building 325, 2800 Kgs. Lyngby, Denmark}

\author[1]{Carsten Heinrich}[]

\author[1]{Kai Heussen}[]

\author[1]{Henrik W. Bindner}[]

\cortext[cor1]{Corresponding author}

\begin{abstract}
The continuous electrification of the mobility and heating sectors adds much-needed flexibility to the power system. 
However, flexibility utilization also introduces new challenges to \glspl{DSO},
who need mechanisms to supervise flexibility activations and monitor their effect on distribution network
operation. Flexibility activations can be broadly categorized to those originating from electricity markets
and those initiated by the \gls{DSO} to avoid constraint violations. Coinciding electricity market driven flexibility
activations may cause voltage quality or temporary overloading issues, and the failure of flexibility activations
initiated by the \gls{DSO} might leave critical grid states unresolved. This work proposes a novel data processing
pipeline for automated real-time identification of fast-ramped flexibility activation events. Its practical value
is twofold: (i) potentially critical flexibility activations originating from electricity markets can be detected
by the \gls{DSO} at an early stage, and (ii) successful activation of \gls{DSO}-requested flexibility can be verified by
the operator. In both cases the increased awareness would allow the \gls{DSO} to take counteractions to avoid
potentially critical grid situations. The proposed pipeline combines techniques from unsupervised detection
and open-set classification. For both building blocks feasibility is systematically evaluated and proofed on real
load and flexibility activation data.
\end{abstract}


\maketitle
\glsresetall
\section{Introduction}
\label{sec:Introduction}
%Motivation and Background
Renewable electricity and electrification are key pillars of global efforts to eliminate fossil fuels in the energy supply. 
The European goal of carbon neutrality in $2050$ is reported to require increased shares of renewable energy and continued electrification of the mobility and heating sectors \cite{european2020impact}. 
This trend will further increase uncertainty and volatility in \glspl{DG}.
Thus, active management of \glspl{DG} based on the emerging smart solutions for monitoring, control, and communication is seen as a requirement for \glspl{DSO} \cite{zhao2014review}.
With the improving capability and affordability of \gls{ICT} the implicit or explicit utilization of local consumption flexibility, commonly referred to as demand response, is becoming more attractive. 
By requesting a load deviation of flexible units during a certain time period such as peak hours, referred to as \gls{FA} event, \glspl{DSO} can use local flexibility to avoid or postpone grid reinforcements \cite{spiliotis2016demand}.  
However, applying local end user flexibility for mitigation of potentially critical network states makes grid operation and security partly dependent on the reliability of third parties. 
Thus, real-time detection of \gls{FA} events is desirable for \glspl{DSO} to verify successful activation, and initiate other measures in case of a failure. 
Moreover, \gls{FA} events are not exclusive to \glspl{DSO}, due to activations originating from electricity markets.
On the one hand, balance responsible parties could request and activate flexibility for portfolio optimization. 
On the other hand, controllable heat pumps and electric vehicles may systematically react to price signals with a sudden change of power consumption \cite{muratori2015residential,schey2012first}.
As a result, \glspl{DSO} are not aware of all \gls{FA} events affecting their network.
If not detected at an early stage, such fast-ramped \gls{FA} events could trigger transformer or line protections due to high coincidence \cite{5589940,krause2011dezentrales} or load rebound effects \cite{sperstad2020impact,HEINRICH2020114399,ZIRAS20191407,mishra2013scaling}. 
The resulting disconnection of customers could lead to high social and financial cost.

The increasing deployment of measuring devices, such as \glspl{mPMU} and \glspl{SM}, increase observability of \glspl{DG} and thus provide the data basis for identification of \gls{FA} events.
%Problem statement
However, real-time identification of \gls{FA} events is challenged by different practical problems. 
The infrequent occurrence of \gls{FA} events limits the available data required for implementation of supervised detection methods.
Moreover, the operation of active \glspl{DG} is influenced by a variety of rare or even unseen event classes such as line faults, topology changes or communication failures \cite{9352761,LABRADORRIVAS2020106602}.
Thus, \gls{FA} event identification also requires differentiation between unknown event classes and \gls{FA} events.
This questions the use of widely applied \gls{CS} classifiers that will falsely classify unknown event classes, due to their inability of rejecting these.
Another challenge for real-time \gls{FA} event identification is seen in the central data processing, e.g. via cloud computing.
Already today the integration of \gls{SM} data in real-time power system operation is limited by communication instead of meter-recording capability \cite{kemal2020trade}. 
Upgrading communication networks entails a high economic burden. 
Moreover, long communication paths increase the possibilities for false data injection attacks and other fraudulent modification of data \cite{muller2021threat, dhirender}.
One approach to overcome the drawbacks of central data processing is seen in a distributed event identification architecture based on edge or fog computing \cite{YOUSEFPOUR2019289}.

The described challenges set specific requirements to the approach and implemented techniques for \gls{FA} event identification.
However, a formulation of these requirements is missing which complicates the development of appropriate strategies and methods for identifying \gls{FA} events.

%Contribution/objective
In this work, a novel \gls{EIP} for fast-ramped \gls{FA} events is proposed. 
A schematic overview of the proposed pipeline is depicted in Fig. \ref{fig:schematic_overview_pipeline}.
The data processing pipeline is based on unsupervised detection and \gls{OS} classification algorithms, suitable for application in a distributed event identification architecture. 
The scheme and algorithm selection are based on a thorough requirements analysis.
The core contributions of this work are the systematic selection of processing algorithms and their evaluation on real load and \gls{FA} event data. 

\input{./Graphics/01_EIP_pipeline.tikz}

\subsection{Related work}
\label{sec:Introduction_related_work}
To the extent of the authors knowledge, this is the first work on detection and classification of \gls{FA} events. 
Works on thematically-related topics are presented first, followed by a presentation of methodologically-related works. 
In both cases literature on detection and classification is discussed separately. 

\subsubsection{Thematically-related works}
A frequently studied topic in power system literature is unsupervised anomaly detection in energy time series data. 
To detect anomalies, most works train models predicting normal behaviour.
A data point is declared an anomaly if deviation between model prediction and ground truth exceeds a predefined threshold.
Various models such as variational autoencoder \cite{pereira2018unsupervised}, \gls{HTM} \cite{barua2020hierarchical}, \gls{ARIMA} or long short-term memory  \cite{hollingsworth2018energy} are applied.
None of these works consider \gls{FA} events as anomaly.
Moreover, most works assume anomaly-free training data for learning of the normal behaviour. 
Some works exist on flexibility detection on building or device level, which try to quantify the flexible load potential in load data of individual devices or buildings \cite{neupane2014towards, mocanu2016energy}. 
Although the name suggests similarity, the problem under investigation is different to the present work, as these works actually investigate flexibility identification.

The topic of event classification in \glspl{DG} has been studied intensely \cite{Liu.2020}.
Most works investigate the multi-class \gls{CS} classification problem.
Literature considering \gls{OS} classification in a power system context is rare.
Lazzaretti et al. \cite{lazzaretti2013new} first applied one-class classifiers for automatic oscillography classification under existence of unseen event classes in two different approaches. 
The first one considers a single one-class classifier for modeling the boundary of multiple known classes. 
A separate multi-class classifier is applied to differentiate between the known classes.
In the second approach each class is modelled with a separate one-class classifier. 
In the following years other works considered one-class classifiers for detection of new classes \cite{lazzaretti2016novelty,huang2016mechanical}. 
Although these methods in general can be used for \gls{OS} classification, the different problem setting results in a comparatively low detection performance \cite{Geng.2020,mahdavi2021survey}. 
To the best of the authors' knowledge, the literature provides no work applying classifiers inherently made for the \gls{OS} problem to event classification in active \glspl{DG}. 
With respect to the proposed \gls{EIP}, some works on event classification exist that assume an upstream detection step \cite{Niazazari.2018, Phillips.2014}.
However, none of these works describe how input samples for the classifier are generated based on the detector results, and rather investigate event classification for existing samples.

\subsubsection{Methodologically-related works}
Similar to literature on anomaly detection in energy time series data, multiple works propose forecasting-based unsupervised anomaly detection \cite{Ahmad.2017,yaacob2010arima,ergen2019unsupervised}. 
In most cases, the euclidean distance between point forecast and ground truth is used to flag anomalies based on a defined threshold. 
In \cite{Munir.2019}, the authors propose the use of a \gls{CNN} as the time series forecaster. 
According to the authors, the proposed method can be trained on comparatively few training data and without removing anomalies from the training dataset. 
A novel approach on anomaly detection in time series data is proposed in \cite{Ren.2019}.
The authors introduce the use of the \gls{SR} algorithm from saliency detection in computer vision for unsupervised anomaly detection in time series. 

In contrast to the traditional \gls{CS} classification problem, less literature exists on \gls{OS} classification.
Scheirer et al. \cite{Scheirer.2013} first formalized the \gls{OS} classification problem and proposed the 1-vs-Set machine as a preliminary solution. 
Since then various methods such as distance-based \cite{bendale2015towards}, margin distribution\hspace{0pt}-based \cite{Rudd.2018} or generation-based \cite{CD_OSR} \gls{OS} classifiers were proposed.
In \cite{Geng.2020}, the authors provide a systematic categorization of \gls{OS} classification techniques, and compare a number \gls{OS} classifiers on popular benchmark datasets.

\subsection{Contribution and paper structure}
The main contributions of this work are as follows:
\begin{itemize}
    \item Proposal of a novel \gls{EIP} based on unsupervised detection and \gls{OS} classification.
    \item First work on detection and classification of \gls{FA} events.
    \item First application of an \gls{OS} classifier to events in active \glspl{DG}.
    \item Introduction of a new performance metric for the evaluation of real-time detection of \gls{FA} events.
    \item Systematic demonstration and evaluation of unsupervised detection and \gls{OS} classification of fast-\hspace{0pt}ramped \gls{FA} events as the main building blocks of the proposed pipeline based on real load and \gls{FA} event data.
\end{itemize}

The remainder of the paper is structured as follows: In Section \ref{sec:Requirements} requirements for \gls{FA} event identification in active \glspl{DG} are evaluated and strategies are proposed. 
Section \ref{sec:Modeling} provides a description of models and methods. 
In Section \ref{sec:Experimental_setup} the experimental setup is presented, including the dataset under investigation, data preparation for model development and evaluation, and applied performance metrics. 
In Section \ref{sec:Results} results are presented and discussed followed by a conclusion and a view on future work in Section \ref{sec:Conclusion}.
\section{Requirements and strategies for \gls{FA} event identification} 
\label{sec:Requirements}

Identifying \gls{FA} events in active \glspl{DG} comes with specific requirements not only concerning the general approach, but also the implemented algorithms.
Additional requirements at algorithmic level are introduced by the consideration of event identification based on a distributed architecture.
In the following, the identified requirements are presented.
Based on the requirements analysis, the concept of the proposed \gls{EIP} as well as the selection of specific models for the main building blocks of the pipeline are justified.
A detailed explanation of the implemented models and the proposed pipeline follows in Section \ref{sec:Modeling}.
The problem is limited to fast-ramped load reduction and load increase \gls{FA} events with a length of up to $3$ hours.
Moreover, the aggregated active power load profile is assumed to represent averaged active power measurements on secondary substation level or aggregated \gls{SM} data collected in a data hub on neighborhood level.
In both scenarios a large low-voltage feeder is considered.
The core of the concept is the separation of the event identification task into an unsupervised event detection and a supervised classification task, resulting in the proposed \gls{EIP}.
Besides the two main building blocks, an event sampler is required to prepare event observations for the classifier based on the results of the event detector.

\subsection{Real-time identification}
A key requirement for identification of \gls{FA} events is real-time capability.
Real-time identification allows \glspl{DSO} to take immediate counteractions in cases where \glspl{FA} could result in critical situations, such as congestions and under or over-voltages.
Supervised detection or classification of time series events usually requires as input the entire time series sample \cite{Gupta.2020}.
For real-time event identification this becomes a fundamental problem.
Existing early classification techniques come at the cost of decreased accuracy \cite{Gupta.2020}, and are not applicable to an \gls{OS} classification problem.
In the proposed pipeline the problem of prediction delay is addressed by separating event identification into two consecutive steps. 
The use of an unsupervised, point-wise event detector allows for immediate flagging of abnormal data points in real-time. 
Although this cannot solve the intrinsic problem of supervised classification being dependent on multiple data points of an event, information extraction is improved.
Instead of identifying an event at the end of its occurrence, with the presented \gls{EIP} \glspl{DSO} will immediately be aware of the existence of a deviation from normal operation, followed by an ex-post classification of the event. 

\subsection{Model development based on limited and partly-labeled training data}
\glspl{FA} in active \glspl{DG} constitute rare events. 
Thus, comprehensive datasets of \gls{FA} events for supervised methods are difficult to obtain.
The heterogeneity of \glspl{DG} and flexibility portfolios brings additional challenges for acquiring datasets, since characteristics of \glspl{FA} will differ for different networks.
In contrast, unsupervised event detection does not require datasets of \gls{FA} events.
Instead, most works on unsupervised event detection in energy time series data, such as \cite{pereira2018unsupervised}, assume event-free training data to learn a representation of the normal behaviour.
However, existing training data will most likely contain events, since manual removing is a time-consuming and impractical process \cite{Ziras.2021} and \glspl{DSO} might not be aware of all \glspl{FA} (see Section \ref{sec:Introduction}).
In the proposed pipeline a persistence forecast-based detector is considered, which is not dependant on event-free training data. 
By applying an unsupervised detector with no demand for event-free training data the dependency on labeled training data is reduced to the classification step.
Therefore, compared to an one-step event identification approach, the proposed pipeline maintains event detection capability also in scenarios without labeled training data, maximizing information extraction. 

\subsection{Lightweight models for event identification} \label{sec:lightweight}
As described in Section \ref{sec:Introduction}, distributed event identification in an edge or fog computing scheme requires models and methods to be lightweight. 
The vast number and limited processing power of edge devices as well as the continuously growing amount of data sets time and resource constraints to the development, operation and maintenance of models and methods. 
Therefore, a key requirement for \gls{FA} event identification is seen in keeping computational and maintenance efforts, such as periodical re-training, at a minimum. 
This is considered for the proposed concept in several ways.
First of all, the proposed pipeline entirely works with delta encoded data. 
Delta encoding is a technique for data compression based on differencing sequential data, reducing data communication and storage load \cite{suel2019delta}.
By working with differenced data the proposed pipeline can directly be applied in a system which uses delta encoding for data compression.
Both the detection and classification model within the pipeline only retrieve features from the univariate load time series.
No additional information such as weather or market price data are considered, reducing the requirement for data communication and the dimensionality of the detection and classification problem.
The use of a simple persistence forecast keeps size and computational effort of the proposed detector at a minimum, and avoids the need for frequent re-training.
In the proposed pipeline the classifier only gets activated if the detector has detected a deviation from normal behaviour above a predefined threshold.
This event-triggered scheme avoids continuous running of the classifier which reduces the required processing power.

For \gls{OS} classification the \gls{EVM} model is selected as it comes with several features, making it a comparatively lightweight classifier \cite{Rudd.2018}.
\gls{EVM} is capable of incremental learning which allows for efficient model updating without time and computation intensive re-training. 
Moreover, the model reduction strategy of \gls{EVM} discards redundant data points within a class of training points, allowing for limitation of model size and classification time as dataset size increases. 

\subsection{Handling multiple and unknown event classes}
In active \glspl{DG} a large variety of events with various backgrounds, such as faults and switching actions, can occur.
While in this work the detection performance is evaluated on the basis of fast-ramped \gls{FA} events, in principle an \textit{unsupervised} detector also allows for detection of other fast-ramped events.

Although traditional \gls{CS} classifiers can differentiate between multiple known event classes, introducing new unknown classes will lower classification performance drastically, since observations of unknown classes are wrongly assigned to one of the classes the classifier was trained on \cite{Geng.2020}.
Given that many event classes only occur rarely and new ones might emerge, e.g. due to changes in grid topology, assuming that training data includes sufficient observations to describe all existing events is considered an unrealistic assumption.
An important requirement for \gls{FA} event identification is therefore seen in the capability to differentiate between \gls{FA} events and other event classes by either recognizing known or rejecting unknown event classes.
For that purpose, an \gls{OS} classifier is specifically selected.

\subsection{Extension to new event classes}
For many other events, such as high-impedance faults or sensor failures, real-time identification would add additional value to \glspl{DSO}.
However, adding additional identification models for every event would again violate the aforementioned time and resource constraints.
For this reason, a central requirement is seen in the capability of a model to be extended to identification of additional events while respecting computational and maintenance effort limitations.
The use of an unsupervised event detector allows for the detection of other fast-ramped events beyond the considered \gls{FA} events.
To extend the event identification problem to slow-ramped events, the persistence forecast-based detector needs to be replaced or extended. 
However, due to the modular fashion of the proposed pipeline an extension to slow-ramped events can be achieved without affecting the subsequent classification step. 
With regard to the classification step, the implementation of an \gls{OS} classifier with rejection and incremental learning capability facilitates the extension to new event classes: indicating observations of unknown classes allows for automated collection, facilitating the manual preparation of new event classes.
Once sufficient observations of a new class are collected, the \gls{EVM} model enables efficient incorporation under an incremental update mechanism. 
The model reduction strategy makes \gls{EVM} a sparse \gls{OS} classifier with limited model size also under extension with new classes.
\section{Model description} \label{sec:Modeling}
This section first formulates the problem of unsupervised event detection and \gls{OS} classification and describes the implemented models. 
Thereafter, the proposed \gls{EIP} is explained. 

\subsection{Unsupervised detection of \gls{FA} events} \label{sec:Modeling_unsupervised_event_detection}
In this subsection, the unsupervised event detection problem is formulated. Subsequently, the implemented models for unsupervised event detection are described, namely \gls{HTM}, \gls{ARIMA}, \gls{CNN}, \gls{SR} and Persistence detector. 

\subsubsection{Problem formulation}
The unsupervised detection of \gls{FA} events is formulated as a point anomaly detection problem in univariate time series data.
A point anomaly is considered a data point that significantly deviates from its expected value.
Given a univariate time series $\bm{X} = \{x_{1},x_{2},...,x_{N} \ | \ x_{i} \in \mathbb{R} \forall i\}$, a data point $x_t$ at time $t$ is declared an anomaly if the anomaly score $s_t$, defined as the distance to the expected value $\Hat{x}_t$, exceeds a predefined threshold $\tau$:
\begin{equation}
    s_t = |x_t - \hat{x}_t| > \tau   \label{eq:anomaly_score}
\end{equation}

Although all detectors within this work follow different strategies to calculate $s_t$, they are all either explicitly or implicitly based on fitting a model to the normal behaviour. 
Given the univariate time series $\bm{X}$, all detection models either aim at learning a mapping function $\Phi$ from historical time steps to the next time step 
\begin{equation}
    \hat{x}_t = \Phi([x_{t-w},...,x_{t-1}]), \label{eq:mapping_function}
\end{equation}
or a direct mapping function $\Theta$ from historical time steps to the anomaly score of the next time step 
\begin{equation}
    s_t = \Theta([x_{t-w},...,x_{t-1}]), \label{eq:mapping_score}
\end{equation}
where $w$ is the size of the history window, which can vary for the different detectors.

\subsubsection{\gls{HTM} detector} \label{sec:HTM}
\gls{HTM} is a machine learning technique that is based on the structural and algorithmic properties of the neocortex \cite{hawkins2004intelligence}. 
Compared to many other methods, \gls{HTM} comes with several advantages that simplify handling of the anomaly detection problem.
These include continuous online learning capability, robustness to noise, and applicability without case-specific hyperparameter tuning. 
Thus, \gls{HTM} can be applied without model training and selection on separate training and validation datasets and frequent re-training. 
For a detailed description of \gls{HTM} the reader is referred to \cite{wu2018hierarchical}. 

The implementation of \gls{HTM} includes an internal calculation of anomaly scores such that the \gls{HTM} detector overall follows \eqref{eq:mapping_score}.
\gls{HTM} comes with approximately $30$ model configuration parameters.
For the anomaly detection problem, a set of optimal parameters is provided in the supplementary material of \cite{Ahmad.2017}, which is applied in this work. 

\subsubsection{\gls{ARIMA} detector} 
\gls{ARIMA} models are widely known and applied for time series forecasting \cite{DEB2017902}.
As they use lagged values to forecast future behaviour, \gls{ARIMA} models can be applied to learn a mapping function according to \eqref{eq:mapping_function}. 
To enable application to seasonal data, this work considers modeling of seasonal patterns based on Fourier terms as proposed in \cite{hyndman}. 
Model training and selection is based on the first $10$ days of the dataset.
In a pre-processing step, the distribution of the training data is centered on a mean of $\mu = 0$ and a standard deviation of $\sigma = 1$.
Based on the \gls{AIC} a ($p,d,q) = (8,1,2$) \gls{ARIMA} model is chosen.
After the initial training, the autoregressive and moving average parameters are updated with every new incoming observation.
Every $14$ days an entirely new \gls{ARIMA} model is selected based on the \gls{AIC}, resulting in a new selection of $p$, $d$ and $q$. 

\subsubsection{\gls{CNN} detector}
\glspl{CNN} \cite{o2015introduction} are a specialized class of artificial neural networks that can capture temporal patterns in time series data. Thus, they can be applied to provide a mapping function according to \eqref{eq:mapping_function}. 
An advantage of \glspl{CNN} is the good performance also for small training datasets (as opposed to many other deep learning techniques) even without removing anomalies from the training data \cite{Munir.2019}. 

To define the architecture and hyperparameters of the \gls{CNN}, extensive empirical experiments are conducted based on the first $10$ days of the dataset. 
The model is trained on predicting the difference $x_{\Delta ,t}=x_{t}-x_{t-1}$ instead of $x_{t}$ to avoid learning a local minimum given by the persistence forecast $\hat{x}_{t}=x_{t-1}$.
To enable faster convergence the training data is centered on a mean of $\mu = 0$ and a standard deviation of $\sigma = 1$.
While the first $7$ days are used as an initial training dataset, the remaining $3$ days are used for validation. 
The resulting \gls{CNN} architecture consists of three convolutional/max-pooling pairs followed by a fully connected layer. 
An overview of the main hyperparameters is given in Table \ref{Tab:CNN_hyperparameter}. 
After the initial training and model selection phase the \gls{CNN} is re-trained every $14$ days based on the previous data.
To avoid overfitting, a combination of early stopping, L2 regularization and dropout is applied. 
\begin{table}[width=1\linewidth]
\caption{Summary of hyperparameters of the CNN model.}
\label{Tab:CNN_hyperparameter}
\centering
\begin{tabular*}{\tblwidth}{@{} LLL@{} }
\toprule
Hyperparameter & Search space & Selected value \\
\midrule
History window size & $144$, $288$, $576$, $1152$ & $288$ \\
Forecasting horizon & $1$ & $1$ \\
Learning rate & [$0.00001$, $0.1$] &$1$e-$5$ \\
L2 weight regularization & [$0.0001$, $0.1$] & $0.01$\\
Dropout rate & [$0$, $0.2$] & $0.2$\\
Batch size & $10$, $50$, $100$, $500$ & $50$\\
\makecell[l]{Maximum number \\ of epochs} & $2000$ & $2000$\\
Number of filters & $10$, $30$, $50$, $70$ & $50$ \\
Kernel size & $2$, $3$, $4$ & $3$ \\
\makecell[l]{Neurons in the fully \\ connected layer} & $10$, $50$, $100$, $150$ & $100$ \\
Early stopping patience & $10$, $50$, $100$ & $50$ \\
Activation function & ReLU, Tanh & ReLU \\
\bottomrule
\end{tabular*}
\end{table}

\subsubsection{\gls{SR} detector}
\gls{SR} \cite{hou2007saliency} is an unsupervised algorithm for visual saliency detection in computer vision. Recently, Ren et al. \cite{Ren.2019} proposed the use of \gls{SR} for anomaly detection in time series data, motivated by the similarity to visual saliency detection. 
An advantage of \gls{SR} is the comparatively small number of hyperparameters.
\gls{SR} performs a mapping of previous data points to the next time step according to \eqref{eq:mapping_function}.
However, the mapping is conducted within a saliency map representation of a sliding sequence $\bm{x} = \{x_{1},x_{2},...,x_{N}\}$, which is based on Fourier transformation and calculation of the spectral residual.
Within the saliency map a local average of previous data points is compared to the actual value to declare anomalies similar to \eqref{eq:anomaly_score}.
As this work is concerned with real-time detection, only the most recent data point $x_{N}$ of sequence $\bm{x}$ is evaluated. 
Since detection performance of \gls{SR} improves for data points located in the center of $\bm{x}$, Ren et al. \cite{Ren.2019} propose to add estimated data points following $x_{N}$.
In this work, the hyperparameters are selected according to the selection in \cite{Ren.2019}, which results from an empirical investigation of multiple datasets for time-series anomaly detection. 
The number of estimated points is set to $5$, the size for the local average is $21$ and the length of the sequence $\bm{x}$ is set as $1440$. 
It is worth noting that a parameter study on the dataset under investigation revealed low sensitivity of the \gls{SR} detection performance to the selection of hyperparameters. 

\subsubsection{Persistence detector} 
In this work, the use of a persistence forecast for modeling the expected value $\hat{x}_t$ is proposed and compared to the previously introduced methods. 
The proposed Persistence detector considers a history window $w = 1$ and determines $\hat{x}_t$ according to
\begin{equation}
    \hat{x}_t = x_{t-1}.
\end{equation}

The triviality of the Persistence detector eliminates the need for model selection, training and re-training. 

\subsection{\gls{OS} classification of \gls{FA} events} \label{sec:Open_set_classification}
This subsection first formulates the \gls{OS} classification problem.
Thereafter, the implemented \gls{OS} classifier is described. 

\subsubsection{Problem formulation} \label{sec:OS_problem_formulation}
\begin{figure*}
\centering
\input{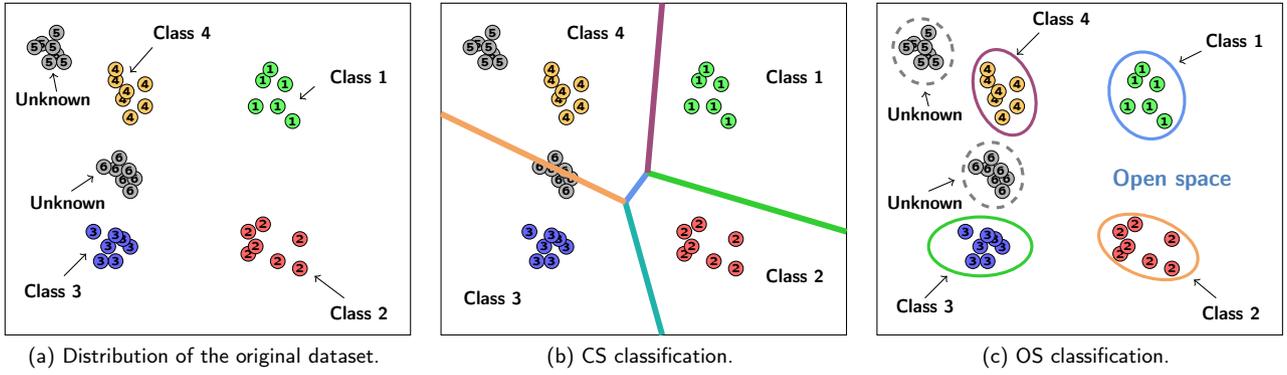}
\caption{Comparison of \gls{CS} and \gls{OS} classification according to \cite{Geng.2020}.}
\label{fig:open_set_problem}
\end{figure*}

\gls{OS} classification is contrasted with \gls{CS} methods typically applied in the literature. 
\gls{CS} classification requires identical event classes in training and test data and thus assumes full awareness of all existing event classes.
For example, a \gls{CS} classifier trained on observations of line failures and tap change operations will declare every observation of an unknown event class as either a line failure or a tap change.
For \glspl{DSO} it might be impractical to obtain data comprising observations off all existing classes, since events such as failures of transmission elements, large loads or sensors rarely occur.
An \gls{OS} classifier rejects observations of event classes not previously seen and declares them as $unknown$.
In Fig. \ref{fig:open_set_problem} \gls{CS} classification is compared to \gls{OS} classification.
Making \gls{CS} assumptions leads to regions of unbounded support, as can be seen from Fig. \ref{fig:open_set_problem} (b). 
This will result in missclassification of observations from unknown classes which can drastically weaken the performance of the classification.

The problem of \gls{OS} classification can be formulated as follows \cite{Yu.24052017}. 
Let $\bm{D}_{\text{train}} = \{(\bm{v}_i, y_i)\}^{N_{\text{train}}}_{i=1}$ be a training dataset, with $\bm{v}_{i} \in\mathbb{R}^{d}$ being a feature vector instance and $y_{i} \in \bm{Y}_{\text{train}} = \{1,2,...,K\}$ the corresponding event class label.
During test or application a classifier needs to predict event classes of the open dataset $\bm{D}_o = \{(\bm{v}_i, y_i)\}^{\infty}_{i=1}$, where $y_i \in \bm{Y}_o = \{1,2,...,K,K+1,...,M\}$ with $M>K$. The occurrence of unknown classes requires the classifier to learn a mapping function $f(v): \bm{V} \rightarrow \bm{Y}^{'} = \{1,2,...,K,unknown\}$, with the option $unknown$ representing the rejection of classes not seen during training.

Similar to the described situation of a \gls{DSO} not having comprehensive recordings of event classes, in the present dataset the number of \gls{FA} events is comparatively small.
Thus, the size of the input feature vectors $\bm{v}_{i}$ is limited to $6$. 
Reducing the dimension of the feature space that needs to be described by the limited number of training observations allows for better determination of the decision boundaries. All features are derived from the delta encoded time series (see Subsection \ref{sec:EIP}) and are listed in Table \ref{Tab:EVM_features}. The definition of the input features is given based on a delta encoded sequence observation $\bm{x}_{\Delta} = \{x_{\Delta,1},...,x_{\Delta,N_{x}}\}$.
\begin{table}[width=1\linewidth,cols=2,pos=ht]
\caption{Overview of features used for \gls{OS} classification of \gls{FA} events.}
\label{Tab:EVM_features}
\centering
\begin{tabular*}{\tblwidth}{@{} LL@{} }
\toprule
Feature & Definition \\
\midrule
Mean $\mu_{\bm{x}_{\Delta}}$ & $\frac{1}{N_{\bm{x}}}\left(\sum^{N_{\bm{x}}}_{i=1}x_{\Delta,i}\right)$\\
Standard deviation $\sigma_{\bm{x}_{\Delta}}$ & $\sqrt{\frac{1}{N_{\bm{x}}-1}\sum^{N_{\bm{x}}}_{i=1}(x_{\Delta,i}-\mu_{\bm{x}_{\Delta}})^2}$\\
Minimum value $x_{\Delta,\text{min}} $ & $\min(\bm{x}_{\Delta})$\\
Maximum value $x_{\Delta,\text{max}} $ & $\max(\bm{x}_{\Delta})$\\
Number of zeros $n_{0}$ & $\text{count}(\bm{x}_{\Delta}\stackrel{!}{=}0) $  \\ 
\makecell[l]{Points between minimum and \\ maximum value $n_{\text{minmax}}$} & $|\text{index}(x_{\Delta,\text{min}})-\text{index}(x_{\Delta,\text{max}})|$ \\
\bottomrule
\end{tabular*}
\end{table}

\subsubsection{\gls{EVM} classifier} \label{sec:EVM}
The \gls{EVM} is an \gls{OS} classifier proposed by Rudd et al. \cite{Rudd.2018}. 
Advantages of the \gls{EVM} include incremental learning and model reduction capability, which avoids frequent re-training and keeps model size and classification time small. 
The \gls{EVM} models known classes within the training dataset by a set of radial inclusion functions (see Fig. \ref{fig:open_set_problem}  (c)).
Based on the radial inclusion function of a class $C_{l}$, the probability $\hat{P}(C_{l}|v')$ of a new observation $v'$ belonging to $C_{l}$ can be determined.
The decision function of the \gls{EVM} is given by 
\begin{equation}
    y^{*} =   
    \begin{cases}
    \text{arg}\,\max_{l \in \{1,...,M\}}  \hat{P}(C_{l}|v')& \text{if} \hat{P}(C_{l}|v') \geq \rho  \\
    unknown & \text{otherwise} \\
  \end{cases},
\end{equation}
where $\rho$ is a threshold defining the boundary between the set of known classes and the unknown open space.

The \gls{EVM} model training and selection is based on $\SI{90}{\percent}$ of the available event observations applying $5$-fold time series cross-\hspace{0pt}validation. The features are standardized based on training data for every individual split. In order to select an appropriate threshold $\rho$, a minimum performance requirement on the training dataset is defined based on the $F_{\text{1}}$ score (Subsection \ref{sec:performance_metrics}). The model with the smallest threshold $\rho$, which still fulfills the performance requirement $F_{1} \geq 0.8$ in the time series cross-\hspace{0pt}validation, is selected. An overview of the selected hyperparameters is given in Table \ref{Tab:EVM_hyperparameter}.

\begin{table}[width=1\linewidth,cols=3]
\caption{Summary of hyperparameters of the \gls{EVM} model.}
\label{Tab:EVM_hyperparameter}
\centering
\begin{tabular*}{\tblwidth}{@{} LLL@{} }
\toprule
Hyperparameter & Search space & Selected value \\
\midrule
Tailsize & [$1$,$100$]  & $7$ \\
\makecell[l]{Distance \\ multiplier} &  [$0.1$,$1.1$] & $0.9$ \\
Distance metric & \makecell[l]{ Canberra distance, \\ Cosine distance, \\Euclidean distance } & Canberra distance \\
Threshold & [$0.1$, $0.99999$] & $0.9$    \\
\bottomrule
\end{tabular*}
\end{table}

\subsection{\gls{EIP}} \label{sec:EIP}
To connect the presented models for unsupervised event detection and \gls{OS} classification, a data processing pipeline is proposed.
In Fig. \ref{fig:schematic_overview_pipeline} a schematic overview of the \gls{EIP} is depicted.
The functionality of the main building blocks of the pipeline was described in Subsection \ref{sec:Modeling_unsupervised_event_detection} and \ref{sec:Open_set_classification}.
In Subsection \ref{sec:lightweight} the use of delta encoded data for the \gls{EIP} was motivated.
Delta encoding exploits the autocorrelation of time series data.
In the simplest version of delta encoding, a time series $\bm{X} = \{x_{1},x_{2},...,x_{N}\}$ is encoded as difference between successive samples, resulting in the delta encoded time series $\bm{X}_{\Delta} = \{x_{1},x_{2}-x_{1},...,x_{N}-x_{N-1}\} =  \{x_{1},x_{\Delta,2},...,x_{\Delta,N}\}$. 
Delta encoding performs best when the values in the original data contain only small changes between adjacent values \cite{smith1997scientist}.
By applying the Persistence detector on the delta encoded time series $\bm{X}_{\Delta}$, the anomaly detection problem reduces to comparison of the amplitudes of $x_{\Delta,i}$ to the predefined threshold $\tau$.
To connect point-wise unsupervised event detection with \gls{OS} classification, an event sampler is interposed, exploiting the characteristics of fast-ramped events. 
As can be seen in Fig. \ref{fig:example_of_different_event_classes} fast-ramped \gls{FA} events show peaks in the delta encoded data at the beginning and/or end of an event. 
Based on this property, the detection of an event at data point $x_{\Delta,t}$ can be used to extract a backward sequence sample $\bm{x}_{\Delta,\text{bw}}=\{x_{\Delta,t-w_{\bm{x}}},...,x_{\Delta,t+e}\}$ and forward sequence sample $\bm{x}_{\Delta,\text{fw}}=\{x_{\Delta,t-e},...,x_{\Delta,t+w_{\bm{x}}}\}$, where $w_{\bm{x}}$ is the sample window size and $e$ a window extension, ensuring sampling of the entire event.
In case of forward sampling, an early stopping criterion can be introduced which breaks the sampling process in case another event is detected at a data point $x_{\Delta,t+a} \in \{x_{\Delta,t+1},...,x_{\Delta,t+w_{\bm{x}}}\}$, resulting in a forward sample $\bm{x}_{\Delta,\text{fw}}=\{x_{\Delta,t-e},...,x_{\Delta,t+a+e}\}$. 
Such event-triggered early stopping of the forward sampling process reduces the sampling time and thus the time until a sample can be classified.
In the Appendix in Algorithm \ref{algo:pipeline} the general procedure of the proposed \gls{EIP} is described.
\section{Experimental setup} \label{sec:Experimental_setup}
This section presents the experimental setup of this study.
The considered dataset as well as the preparation of the dataset for investigation of unsupervised event detection and \gls{OS} classification of \gls{FA} events are presented in Subsection \ref{sec:Dataset}. 
Subsection \ref{sec:performance_metrics} introduces metrics for the evaluation of the detection and classification performance. 

\subsection{Dataset and data preparation} \label{sec:Dataset}
The first part of this subsection is concerned with presenting key information of the dataset under investigation as well as describing the process of \gls{FA}. 
In the second part the preparation of the dataset is explained, which includes data cleaning and in case of preparation for the investigation of the \gls{OS} classification the extension of the dataset with two artificial event classes. 

\subsubsection{Dataset}
\label{sec:dataset_sub}
\begin{figure*}
\centering
    \input{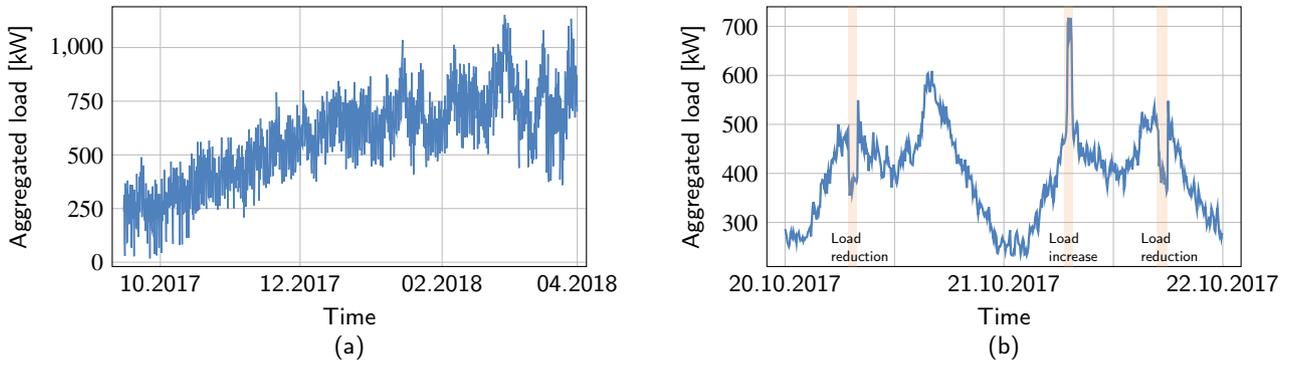}
    \caption{Aggregated load dataset (a) and aggregated load of two representative days with frequent \glspl{FA} (b). Periods of \glspl{FA} are marked with an orange background.} \label{fig:Dataset_entire_and_week}
\end{figure*}

Within this work, a dataset from EcoGrid~2.0 is used.
EcoGrid~2.0 was a demonstration project which examined the use of flexible consumption of residential customers for power system services at transmission system operator and \gls{DSO} level \cite{HEINRICH2020114399}. 
The experiments were conducted on the Danish island of Bornholm.
The residential customers were eq-\\uipped with \glspl{SM} and \gls{ICT} infrastructure for participating in demand response experiments.
The flexible load corresponded to electric heaters and heat pumps that were controlled by adjusting room temperature setpoints or by sending a throttle signal, respectively.
An increase in the setpoints results in higher consumption, while lowering the setpoints leads to a reduction of consumption
The throttle signal blocks the operation of the heat pump until the signal is released.
Besides flexible load, the installed \glspl{SM} also capture household consumption and photovoltaic production, when present.
As can be seen in Fig. \ref{fig:Dataset_entire_and_week} (a), the dataset used consists of six and a half months of aggregated load data, beginning from 15\textsuperscript{th} of September 2017.
The aggregated active power profile consists of $450$ household loads with a 5 minute time resolution.
The \gls{FA} events consist of load reduction and load increase experiments (see Fig. \ref{fig:Dataset_entire_and_week} (b)) realized by two different aggregators and customer portfolios.
Activation periods are in the range of $30$-$120$ minutes. 
Different numbers of customer loads participated in each \gls{FA}, and activations were conducted under varying conditions of temperature, time of the day and photovoltaic production.
In Fig. \ref{fig:Dataset_entire_and_week} (a) the trend and (b) seasonality of the non-stationary load time series can be noticed.

\subsubsection{Dataset preparation for investigation of unsupervised event detection}
The dataset contains $325$ \gls{FA} events.
Start and end time as well as type of \gls{FA} event is known for every experiment.
In $75$ cases two flexibility portfolios were activated simultaneously. 
To avoid double counting of either \glspl{TP} or \glspl{FN}, parallel experiments are considered as one \gls{FA} event, reducing the number of \gls{FA} events to $250$.
In most cases, experiments result in either load reduction or load increase of a subset of customer loads.
However, in some cases little to no flexibility was activated.
This can be due to exclusive testing of connectivity, failed activation of flexibility assets, or low flexibility potential due to high temperatures and thus low heating demand.
For this work, such \gls{FA} events are not considered in the performance evaluation, reducing the number of \gls{FA} event samples to $205$. 

\subsubsection{Dataset preparation for investigation of \gls{OS} event classification}
\begin{figure*}[ht]
\centering
\input{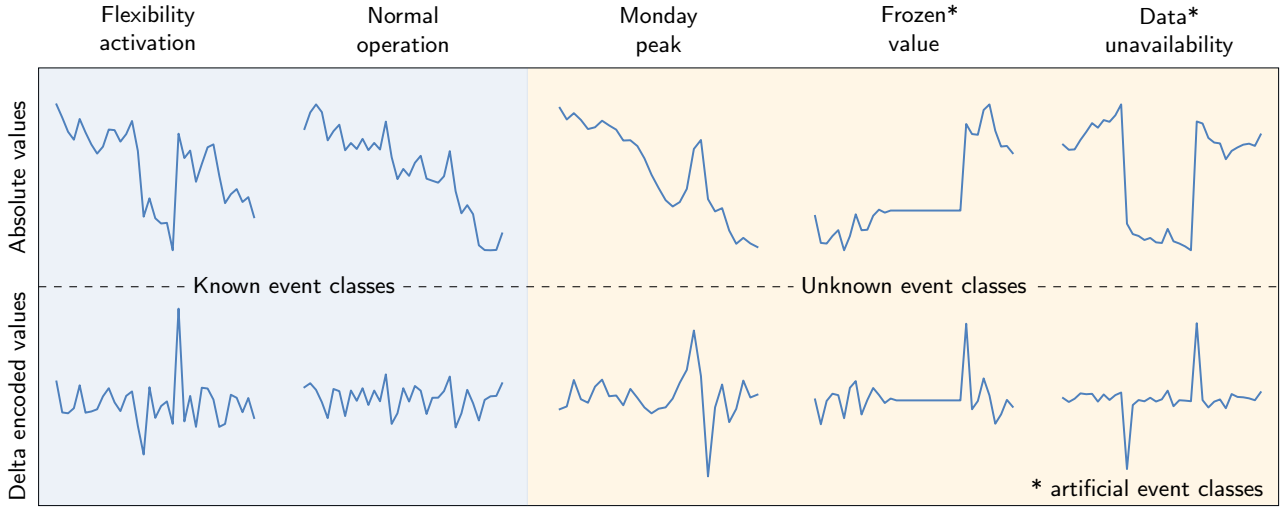}
\caption{Exemplary representation of event classes considered in this work.}
\label{fig:example_of_different_event_classes}
\end{figure*}

In Fig. \ref{fig:example_of_different_event_classes} examples of all event classes are depicted in absolute and delta encoded values. 
Note that only for the \gls{OS} classification problem all introduced event classes are considered.
For the \gls{FA} event detection problem only \gls{FA} events are taken into account. 
The classifier is trained on two known event classes, namely \gls{FA} and \gls{NO} events.
All $205$ \gls{FA} events are sampled, including $3$ timestamps ($15$ minutes) that precede and succeed each event.
As the duration of \gls{FA} events varies, the length of \gls{FA} event samples varies as well. 
An equal amount of \gls{NO} events are randomly sampled from the remaining dataset.
The length of a \gls{NO} event is randomly selected from the distribution of \gls{FA} event lengths. 
With this approach, the classifier is prevented from differentiating between \gls{FA} and \gls{NO} events based on the sample length. 
As the sample length of both \gls{FA} and \gls{NO} events can vary in the proposed \gls{EIP} (see Subsection \ref{sec:EIP}), learning a constant sample length is not considered a valid approach.

The problem of \gls{OS} classification, as formulated in Subsection \ref{sec:OS_problem_formulation}, requires additional event classes within the test dataset to investigate the capability of rejecting unknown classes.
In this work, $3$ unknown event classes are considered. 
Besides the \gls{FA} and \gls{NO} event classes, the EcoGrid 2.0 dataset includes another event class, which in the course of this work will be called \gls{MP}. 
On every Monday within the dataset, a load peak occurs at around $8$ am.
The load peak results from short, collective heating of electric water boilers to \SI{80}{\degreeCelsius} to inhibit the growth of bacteria.  
According to the \gls{FA} events, only \gls{MP} events that could be manually detected in an extensive ex-post evaluation of the dataset are considered. 
\gls{MP} events are not considered a normal operation and thus no overlapping of \gls{NO} and \gls{MP} events exists.
In total the dataset includes $15$ \gls{MP} events. 
In order to extend the \gls{OS} classification problem, two additional artificial event classes are introduced, namely the \gls{FV} and \gls{DU} event class.
The \gls{FV} event class models a data transmission or processing failure in which a measurement at time $t_{0}$ remains constant for $N_{\text{cons}}$ consecutive steps. 
At $t_{N_{\text{cons}}+1}$ recording of true measurements is reestablished.
The length of \gls{FV} events is randomly selected from the distribution of \gls{FA} event lengths.
$205$ \gls{FV} events are randomly introduced into the subset of the dataset which is not influenced by \gls{FA}, \gls{NO} and \gls{MP} events.
In a \gls{DU} event a subset of individual measurements, e.g. \gls{SM} readings, is considered to be unavailable due to device or data transmission failure. 
The fraction of available measurements is randomly selected from the uniform distribution $\mathcal{U}(0.4, 0.8)$. 
As for \gls{NO} and \gls{FV} events, the length of \gls{DU} events is drawn from the distribution of \gls{FA} event lengths.
$205$ \gls{DU} events are introduced into the subset of the dataset not affected by any of the previously described events. 

\subsection{Performance metrics} \label{sec:performance_metrics}
The performance evaluation of both the unsupervised detection and \gls{OS} classification of \gls{FA} events is based on a labeling of each instance of the respective dataset. 
The definition of an instance for the detection and classification task follows in Sections \ref{sec:metrics_detection} and \ref{sec:metrics_classification}, respectively.
One performance metric used for evaluation of both the detection and classification part is the ${F}_{1}$ score. 
The ${F}_{1}$ score represents the harmonic mean between precision and recall, and is a widely applied performance metric for detection and classification problems with imbalanced classes.
Let $\mathit{TP}_{l}$, $\mathit{FP}_{l}$, $\mathit{TN}_{l}$, and $\mathit{FN}_{l}$ be the number of \glspl{TP}, \glspl{FP}, \glspl{TN}, and \glspl{FN} for the $l$th event class, respectively, where $l \in \{1,2, \ldots ,M\}$.
For a multi-class problem, the macro ${F}_{1}$ score is calculated by
\begin{equation}
%% \authornote
%% We accidently included the definition of the micro F1 score instead of macro F1 score. Thus, respective equations were updated.
%% \endauthornote
    F_{1} = \frac{1}{M}\sum^{M}_{l=1}F_{1,l}=\frac{1}{M}\sum^{M}_{l=1}2\frac{Pr_{l} \cdot Re_{l}}{Pr_{l}+Re_{l}},  \label{eq:F1}
    %F_{1} = 2  \frac{Pr \cdot Re}{Pr + Re},  \xlabel{eq:F1}
\end{equation}
where precision $Pr$ and recall $Re$ of the $l$th event class are defined as
\begin{equation}
    Pr_l = \frac{\mathit{TP}_{l}}{(\mathit{TP}_{l}+\mathit{FP}_{l})}, \, \text{and} \ Re_l = \frac{\mathit{TP}_{l}}{(\mathit{TP}_{l}+\mathit{FN}_{l})}.  \label{eq:PrRe}
\end{equation}

\subsubsection{Unsupervised event detection metrics} \label{sec:metrics_detection}
In this work, the problem of unsupervised event detection is considered an anomaly detection problem, reducing the number of classes to $M=2$. 
Since anomalies constitute the primary class of interest, the multi-class formulation of the ${F}_{1}$ score in \eqref{eq:F1} reduces to
\begin{equation}
    F_{1} = F_{1,1}=2\frac{Pr_{1} \cdot Re_{1}}{Pr_{1}+Re_{1}}.
    %Pr = \frac{\mathit{TP}_{1}}{\mathit{TP}_{1}+\mathit{FP}_{1}},  \ \text{and} \ Re = \frac{\mathit{TP}_{1}}{\mathit{TP}_{1}+\mathit{FN}_{1}}   
\end{equation}

Besides the $F_{1}$ score, another widely applied performance metric for anomaly detection is the area under the precision-recall curve (\textit{AUCPR}).
Precision-recall curves summarize the trade-off between precision and recall for different thresholds $\tau$.
While the consideration of \glspl{TN} in traditional receiver-\hspace{0pt}oper\-ating-\hspace{0pt}characteristic curves may lead to an overly optimistic view on the performance in case of highly imbalanced classes, \textit{AUCPR} is specifically tailored to problems with imbalanced classes or rare events.

In this work, the entire sequence of an \gls{FA} event, referred to as event window $\bm{\omega}_{\text{FA}}$, is considered for labeling as \gls{TP} or \gls{FN}.
The event window of a \gls{FA} event is defined by the \gls{FA} start time $t_{\text{FA,start}}$ and end time $t_{\text{FA,end}}$. 
Since the dataset under investigation is a real-world dataset it contains some inaccuracies in the event labeling.
In some cases flexibility was activated before the official start time $t_{\text{FA,start}}$. 
Since in these cases an early detection would result in falsely \glspl{FN}, the event window $\bm{\omega}_{\text{FA}}$ is extended by $10$ minutes, such that $\bm{\omega}_{\text{FA}} = \{x_{t_{\text{FA,start}}-2}, ...,x_{t_{\text{FA,end}}}\}$.
The first detection that falls into $\bm{\omega}_{\text{FA}}$ is considered as \gls{TP}, while further detections within the same event window are ignored.
If no point of $\bm{\omega}_{\text{FA}}$ is detected the event label  will be considered a \gls{FN}. 
In most cases, \glspl{FA} result in a subsequent load rebound, which can be considered a deviation from the normal load behaviour outside of the activation period.
While regarding a detection within the rebound area as \gls{TP} would introduce a positive bias to detection performance, considering them as \gls{FP} would result in overly pessimistic performance results, as the detector indeed has detected an anomaly. 
For this reason, a rebound window $\bm{\omega}_{\text{R}}$ is introduced in which detections are ignored and are thus neither considered a \gls{TP} nor \gls{FP}. 
The length of a rebound window is defined as three times the \gls{FA} event length, resulting in $\bm{\omega}_{\text{R}} = \{x_{t_{\text{FA,end}}+1}, ...,x_{t_{\text{FA,end}}+3 \times N_{\text{FA}}}\}$. 
In contrast to the calculation of \glspl{TP} and \glspl{FN}, the calculation of \glspl{FP} and \glspl{TN} is conducted point-wise.
Thus, all detections outside the event and rebound windows are considered \gls{FP}.

To evaluate the early detection capability, the average detection delay $\bar{\delta}_{\text{det}}$ is introduced as the average time between \gls{FA} start time $t_{\text{FA,start}}$ and the first detection time $t_{\text{det}}$ in minutes according to 
\begin{equation}
    \bar{\delta}_{\text{det}} = \frac{1}{N_{\text{det}}}\sum^{N_{\text{det}}}_{i=1}\delta_{\text{det},i} =  \frac{1}{N_{\text{det}}}\sum^{N_{\text{det}}}_{i=1}(t_{\text{det},i} - t_{\text{FA,start},i}), \label{eq:detection_delay}
\end{equation}
where $N_{\text{det}}$ is the number of detected \gls{FA} events and $\delta_{\text{det},i}$ the detection delay of a detected \gls{FA} event. 
Note that the detection delay of detections is assumed to be zero within the subset $\{x_{t_{\text{FA,start}}-2},x_{t_{\text{FA,start}}-1},x_{t_{\text{FA,start}}}\}$. 
 
The use of widely applied performance metrics allows for an easy understanding and comparison of the results to other studies. 
However, in order to take performance requirements of a specific scenario into account an individual performance metric is required.
For this purpose, the flexibility activation detection score (\textit{FAD} score) is proposed. 
In the scenario of real-time detection of \glspl{FA} in active \glspl{DG}, the cost of \gls{FN} is considered to be higher compared to the cost of \gls{FP}. 
While a missed critical \gls{FA} could lead to violation of power or voltage boundaries, a false alarm would result in a moderate additional manual inspection effort.
Moreover, in the proposed pipeline, a \gls{FP} will lead to a sample of normal behavior (\gls{NO} event class) which can be classified as such by the \gls{OS} classifier. 
In this way the classifier relativizes the \gls{FP} of the unsupervised event detector. 
For these reasons, the \textit{FAD} score puts more weight on \glspl{FN} than on \glspl{FP}. 
Further, early detection capability is an important requirement in the considered scenario. 
The earlier a potentially critical \gls{FA} event is detected, the greater the scope for countermeasures. 
On the contrary, detection near the end of a critical \gls{FA} results in almost no benefit.
The \textit{FAD} score takes these considerations into account and expresses the performance for the specific scenario of real-time \gls{FA} event detection in one score.
Besides an easier evaluation and comparison of detection models, the \textit{FAD} score also allows for easy selection of an optimal threshold $\tau$ for unsupervised detection of \gls{FA} events. 
The proposed \textit{FAD} score is constituted by $3$ scoring functions $\zeta_{\text{TP}}$, $\zeta_{\text{FN}}$ and $\zeta_{\text{FP}}$, representing the contribution of \glspl{TP}, \glspl{FN} and \glspl{FP}, respectively:
\begin{equation}
    \text{\textit{FAD}} = \zeta_{\text{TP}} - \zeta_{\text{FN}} - \zeta_{\text{FP}}.
\end{equation}

Let $x_{t_{\text{FA,start}},i}$, $x_{t_{\text{FA,end}},i}$ and $x_{t_{\text{det}},i}$ be the start point, end point and first detection within the event window $\bm{\omega}_{\text{FA},i}$ of the $i$-th \gls{FA} event, respectively. Then $\zeta_{\text{TP}}$ is given by
\begin{equation}
    \zeta_{\text{TP}} = \sum^{N_{\text{det}}}_{i=1}\sigma_{i}(x_{t_{\text{det}},i}),
\end{equation}
where $\sigma_{i}(x_{t_{\text{det}},i})$ is the positive score of one \gls{TP} detection. Between $\sigma_{i}(x_{t_{\text{FA,start}},i}) = \xi$ and $\sigma_{i}(x_{t_{\text{FA,end}},i}) = 0$ the score $\sigma_{i}$ follows a linear declining function, where $\xi$ is the maximum positive score of one \gls{TP} detection. 
For each missed \gls{FA} event a negative score of $\eta$ is considered according to
\begin{equation}
    \zeta_{\text{FN}} = -\eta \cdot \textit{FN}_{1}.
\end{equation}

Scoring function $\zeta_{\text{FP}}$ is represented by a moved negative exponential function given as
\begin{equation}
    \zeta_{\text{FP}} = -\gamma \cdot  \exp\left(\frac{-\textit{FP}_{1}}{\upsilon}\right) - \nu,
\end{equation}
where $\gamma$ is the maximum negative score of a \gls{FP} detection.
As mentioned previously  $\gamma << \xi$. 
Parameters $\upsilon$ and $\nu$ allow for additional adjustments of the score to the dataset. 
The negative exponential decline considers that the first \gls{FP} detections will have a stronger negative impact on performance, while for subsequent \glspl{FP} the additional negative impact is small. 
The final \textit{FAD} score is normalized such that \textit{FAD}$_{\text{norm}} \in [0,1]$, according to 
\begin{equation}
    \text{\textit{FAD}}_{\text{norm}} = \frac{\text{\textit{FAD}}-\text{\textit{FAD}}_{\text{null}}}{\text{\textit{FAD}}_{\text{opt}}-\text{\textit{FAD}}_{\text{null}}},
\end{equation}
where \textit{FAD}$_{\text{null}}$ is the \textit{FAD} score without any detection and \textit{FAD}$_{\text{opt}}$ the \textit{FAD} score under optimal detection. 
In this work, the parameters are selected as $\xi = 1$, $\eta = 1$, $\gamma = 0.05$, $\upsilon = 10000$ and $\nu = 0$.

\subsubsection{\gls{OS} event classification metrics} \label{sec:metrics_classification}
The performance evaluation of the \gls{OS} classification of \gls{FA} events is based on the ${F}_{1}$ score according to \eqref{eq:F1}.
However, for the \gls{OS} problem the number of classes should only be determined by the known classes. 
Considering all of the unknown classes as a single additional class in the test dataset would result in a biased performance result. 
If the problem is treated in the same way as a \gls{CS} scenario, rejected samples of unknown classes would be considered as \glspl{TP} - although no training samples of the unknown classes existed.
Instead, the calculation of the ${F}_{1}$ score is given by
\begin{equation}
    F_{1} = \frac{1}{K}\sum^{K}_{l}F_{1,l}=\frac{1}{K}\sum^{K}_{l}2\frac{Pr_{l} \cdot Re_{l}}{Pr_{l}+Re_{l}},
    %Pr = \frac{\sum^{K}_{l=1}\mathit{TP}_{l}}{\sum^{K}_{l=1}(\mathit{TP}_{l}+\mathit{FP}_{l})}, \ \text{and} \ Re = \frac{\sum^{K}_{l=1}\mathit{TP}_{l}}{\sum^{K}_{l=1}(\mathit{TP}_{l}+\mathit{FN}_{l})}, 
\end{equation}
where $K$ is the number of known classes from the training dataset. 
In Subsection \ref{sec:results_classification} the influence of the number of unknown event classes on the classification performance will be evaluated, which requires the definition of the \textit{openness} of a test dataset.
In \cite{Scheirer.2013} the authors introduce a formal definition of the openness $O$ of a dataset according to
\begin{equation}
   O = 1 - \sqrt{\frac{2 \times |\text{training classes}|}{|\text{testing classes}|+ |\text{target classes}|}},
\end{equation}
with $ O \in [0,1]$. 
Large values for $O$ correspond to a higher number of unknown classes in the dataset, while for the \gls{CS} problem $O=0$.
\section{Results and discussion} \label{sec:Results}
In this section the performance evaluation of the proposed models for unsupervised detection and \gls{OS} classification of \gls{FA} events is presented. 
In Subsection \ref{sec:results_detection} the proposed Persistence detector is compared to various other models, introduced in Subsection \ref{sec:Modeling_unsupervised_event_detection}. 
Subsection \ref{sec:results_classification} investigates the \gls{OS} classification of \gls{FA} events based on the introduced \gls{EVM} model (Subsection \ref{sec:EVM}).
The classification performance is compared to a \gls{CS} classifier benchmark.
The performance evaluation is conducted based on the dataset and performance metrics from Section \ref{sec:Experimental_setup}.
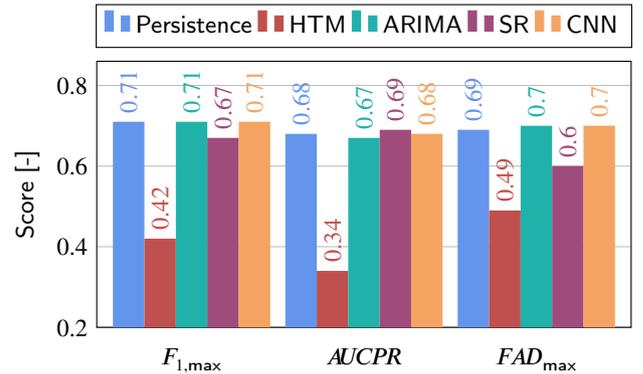
\begin{figure}[b]
\centering
\resizebox{0.48\textwidth}{!}{
\begin{tikzpicture}
    \begin{axis}[
        width  = 0.486\textwidth,
        height = 5cm, 
        major x tick style = transparent,
        major y tick style = transparent,
        ybar=0pt,
        bar width=11.5pt, % 25
        nodes near coords,
        ymajorgrids = true,
        symbolic x coords={$F_{1,\text{max}}$,$A\!U\!C\!P\!R$,$F\!A\!D_{\text{max}}$},
        every node near coord/.append style={rotate=90, anchor=west},
        xtick = data,
        enlarge x limits=0.275,
        ymin=0.2,
        ymax=0.86,
        ylabel = {Score [-]},
        legend cell align=left,
        legend style={
                at={(0.5,1.05)},
                anchor=south,
                column sep=0ex,
                legend columns=5
        }
    ]
        \addplot[style={cornflowerblue,fill=cornflowerblue, draw=none, mark=none}] 
            coordinates {($F_{1,\text{max}}$, .71) ($A\!U\!C\!P\!R$,.68) ($F\!A\!D_{\text{max}}$,.69)};

        \addplot[style={rred,fill=rred,mark=none, draw=none}]
             coordinates {($F_{1,\text{max}}$,.42) ($A\!U\!C\!P\!R$,.34) ($F\!A\!D_{\text{max}}$,.49)};

        \addplot[style={lightseagreen,fill=lightseagreen,mark=none, draw=none}]
             coordinates {($F_{1,\text{max}}$,.71) ($A\!U\!C\!P\!R$,.67) ($F\!A\!D_{\text{max}}$,.7)};
    
         \addplot[style={ppurple,fill=ppurple,mark=none, draw=none}]
            coordinates {($F_{1,\text{max}}$,.67) ($A\!U\!C\!P\!R$,.69) ($F\!A\!D_{\text{max}}$,.6)};
    
         \addplot[style={sandybrown,fill=sandybrown,mark=none, draw=none}]
            coordinates {($F_{1,\text{max}}$,.71) ($A\!U\!C\!P\!R$,.68) ($F\!A\!D_{\text{max}}$,.7)};

        \legend{Persistence,HTM ,ARIMA ,SR , CNN}

    \end{axis}
\end{tikzpicture}}
\caption{Maximum $F_{1}$ score ($F_{1,\text{max}}$), \textit{AUCPR} and maximum \textit{FAD} score (\textit{FAD}$_{\text{max}}$) for Persistence, \gls{HTM}, \gls{ARIMA}, \gls{SR} and \gls{CNN} detector.} \label{fig:detection_barplots}
\end{figure} 
\subsection{Unsupervised \gls{FA} event detection} \label{sec:results_detection}
In Fig. \ref{fig:detection_barplots} the maximum $F_{1}$ score $F_{1,\text{max}}$ and \textit{FAD} score \textit{FAD}$_{\text{max}}$ at the optimal threshold $\tau_{\text{opt,F1}}$ and $\tau_{\text{opt,FAD}}$, respectively, are depicted together with the \textit{AUCPR} for Persistence, \gls{HTM}, \gls{ARIMA}, \gls{SR} and \gls{CNN} detector. 
From the comparison of $F_{1,\text{max}}$ and \textit{AUCPR} it can be derived that Persistence, \gls{ARIMA}, \gls{SR} and \gls{CNN} detector lie in the same performance range.
However, the \gls{HTM} detector shows a significantly poorer performance.
While according to the $F_{1}$ score the Persistence, \gls{ARIMA} and \gls{CNN} detector achieve the best detection results, with $F_{1,\text{max}} = 0.71$, in accordance with the \textit{AUCPR} the \gls{SR} detector outperforms all other detectors with $\text{\textit{AUCPR}} = 0.69$.
Interestingly, with a difference of $4$ percentage points the $F_{1}$ score shows a significant poorer detection performance for the \gls{SR} detector.
Although both the $F_{1}$ score and \textit{AUCPR} are performance metrics specifically tailored to scenarios with highly imbalanced classes and higher emphasis on the positive class, they suggest different results. 
This again motivates the need for a scenario-specific performance metric. 
Based on the comparison of the maximum \textit{FAD} score \textit{FAD}$_{\text{max}}$ in Fig. \ref{fig:detection_barplots} it can be concluded that both the \gls{ARIMA} and \gls{CNN} detector achieve the best result for the problem of real-time detection of \gls{FA} events in aggregated load data with \textit{FAD}$_{\text{max}} = 0.7$. 
On the contrary, with \textit{FAD}$_{\text{max}} = 0.6$ the \gls{SR} detector shows a significant poorer performance in the given case of \gls{FA} event detection. 
However, according to $F_{1,\text{max}}$ and \textit{AUCPR} the \gls{SR} detector can potentially keep up with or even outperform other detection methods in scenarios with other requirements.
With \textit{FAD}$_{\text{max}} = 0.69$ the performance of the Persistence detector is only slightly below the best \textit{FAD} scores achieved by the \gls{ARIMA} and \gls{CNN} detector. 
As can be seen in Fig. \ref{fig:example_of_different_event_classes} \gls{FA} events in the given dataset are fast-ramped events that are characterized by a steep slope at the beginning and end of an event, resulting in a large deviation between $x_{t}$ and $x_{t-1}$. 
In case of the Persistence detector this large deviation directly translates to a large anomaly score according to \eqref{eq:anomaly_score}. 
Given that the Persistence detector can keep up with the more complex detectors regardless of the considered performance metrics, it can be concluded that the Persistence detector constitutes a trivial but effective method for detection of fast-ramped \gls{FA} events.
\begin{figure}[b]
\centering
\input{Graphics/06_Average_detection_delay_tau.tikz}
    \caption{Average detection delay $\bar{\delta}_{\text{det}}$ for Persistence, \gls{HTM}, \gls{ARIMA}, \gls{SR} and \gls{CNN} detector.} \label{fig:detection_detection_delay_line_plot}
\end{figure}

In Fig. \ref{fig:detection_detection_delay_line_plot} the average detection delay $\bar{\delta}_{\text{det}}$ is shown as a function of the threshold $\tau$ for all detectors.
As for $\tau = 0$ all data points are declared an event, the average detection delay is $\bar{\delta}_{\text{det}} = 0$ for all detectors. 
It can be seen that the \gls{HTM} detector has the lowest detection delay for thresholds $\tau > 0.1$. 
The comparatively high early detection capability also explains the reduced performance discrepancy between the \gls{HTM} and the other detectors for \textit{FAD}$_{\text{max}}$ compared to $F_{1,\text{max}}$ and \textit{AUCPR} (Fig. \ref{fig:detection_barplots}). 
While for the \textit{FAD} score the contribution of \glspl{TP} is weighted based on the detection delay, $F_{1,\text{max}}$ and \textit{AUCPR} do not take early detection into account. 

\begin{figure*}
    \centering
    \input{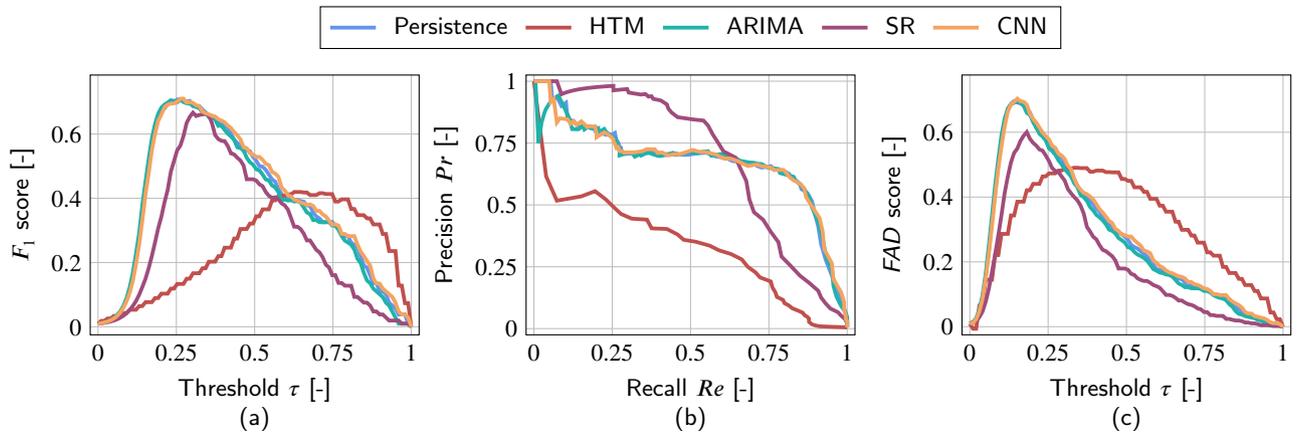}
    \caption{(a): $F_{1}$ score, (b): Precision-recall curve, (c): \textit{FAD} score for Persistence, \gls{HTM}, \gls{ARIMA}, \gls{SR} and \gls{CNN} detector.} \label{fig:F1_score_AUC_PR_line_plot}
\end{figure*}

The Persistence, \gls{ARIMA}, \gls{SR} and \gls{CNN} detector show a similar $\bar{\delta}_{\text{det}}$ for $\tau < 0.4$.
For thresholds $\tau > 0.4$ the \gls{SR} detector shows a significantly higher detection delay, while Persistence, \gls{ARIMA} and \gls{CNN} detector continuously show a similar detection delay. 
\begin{figure}[t]
\centering
\begin{tikzpicture}
    \begin{axis}[
    width  = 1\linewidth,
    height = 5cm,
        major x tick style = transparent,
        major y tick style = transparent,
        nodes near coords,
        ymajorgrids = true,
        %scale only axis,
        separate axis lines,
        xmin=0.25,
        xmax=5.75,
        xtick={1,2,3,4,5},
        x tick style={draw=none},
        xticklabels={Persistence,HTM,ARIMA,SR,CNN},
        ymin=0.5,
        ymax = 13.5,
        ylabel = {$\bar{\delta}_{\text{det}}$ [min]},
        every axis plot/.append style={
          ybar,
          bar width=25,
          bar shift=0pt,
          fill
        }
      ]
        \addplot[style={cornflowerblue,fill=cornflowerblue,mark=none}] 
            coordinates {(1,7.41)};

        \addplot[style={rred,fill=rred,mark=none}]
             coordinates {(2,9.39)};

        \addplot[style={lightseagreen,fill=lightseagreen,mark=none}]
             coordinates {(3,7.49)};
    
         \addplot[style={ppurple,fill=ppurple,mark=none}]
            coordinates {(4,11.42)};
    
         \addplot[style={sandybrown,fill=sandybrown,mark=none}]
            coordinates {(5,8.03)};

    \end{axis}
\end{tikzpicture}
\caption{Average detection delay $\bar{\delta}_{\text{det}}$ at \textit{FAD}$_{\text{max}}$ for Persistence, \gls{HTM}, \gls{ARIMA}, \gls{SR} and \gls{CNN} detector.} \label{fig:detection_detection_delay}
\end{figure}
Fig. \ref{fig:detection_detection_delay} compares the average detection delay $\bar{\delta}_{\text{det}}$ of all detectors at the optimal threshold $\tau_{\text{opt,FAD}}$ corresponding to \textit{FAD}$_{\text{max}}$. 
Although the \gls{HTM} detector has the lowest average detection delay over the largest range of $\tau$ (Fig. \ref{fig:detection_detection_delay_line_plot}), at an operating point relevant for the considered scenario (i.e. at \textit{FAD}$_{\text{max}}$), it has a significantly higher detection delay compared to other detectors.
The \gls{HTM} detector requires a higher threshold compared to the other detectors (see Fig. \ref{fig:F1_score_AUC_PR_line_plot} (c)) , due to the particularly strong vulnerability to high \gls{FP} numbers for low thresholds.
The higher threshold in turn explains the higher average detection delay of the \gls{HTM} detector. 
The \gls{SR} detector with $\bar{\delta}_{\text{det}} = 11.42\,\text{min}$ has by far the highest detection delay.
Persistence, \gls{ARIMA} and \gls{CNN} detector show similar delays. 
In fact, the proposed Persistence detector shows the lowest detection delay with $\bar{\delta}_{\text{det}} = 7.41\,\text{min}$.
However, it has to be considered, that the calculation of the average detection delay is only based on detected events according to \eqref{eq:detection_delay}. 
Thus, detecting additional events close to the end of an event (as done by \gls{ARIMA} and \gls{CNN} detector) results in an improved \textit{FAD} score, as negative \gls{FN} scores are avoided, even though the average detection delay increases.

Fig. \ref{fig:F1_score_AUC_PR_line_plot} (a) and (b) show the $F_{1}$ score over the threshold $\tau$ and the precision-recall curve for the different detectors, respectively. 
Both on the $F_{1}$ score and precision-recall curve a strong similarity of the Persistence, \gls{ARIMA} and \gls{CNN} detector can be noticed.
This can be explained by the signal-to-noise ratio of the dataset.
Although aggregated, the load data under investigation show a comparatively low signal-to-noise ratio due to fluctuations, introduced by unforeseeable customer behavior, and the high resolution of the data.
Because of the low signal-to-noise ratio it is difficult for more complex methods, such as the applied \gls{ARIMA} and \gls{CNN} model, to extract additional information from the dataset compared to the trivial persistence forecast. 
Thus, the explainability of the dataset can be exploited by the persistence forecast to a large extend, explaining the similarity of the Persistence, \gls{ARIMA} and \gls{CNN} detector.
However, the \gls{SR} detector clearly shows a different behavior.
This is due to the different mathematical approach of transforming the dataset from time into the frequency domain.

Fig. \ref{fig:F1_score_AUC_PR_line_plot} (b) shows that, compared to all other detectors, the \gls{SR} detector is able to keep the precision on a higher level for an increasing recall. 
While a high precision is not seen as an important requirement for the considered scenario, the \gls{SR} detector may have advantages over the other detectors in scenarios with different requirements.
Interestingly, the \gls{HTM} detector clearly shows a different behavior compared to the Persistence, \gls{ARIMA} and \gls{CNN} detectors, even though it is also based on a time series forecast. 
This can partly be explained by the internal calculation of the anomaly score, which differs from the external calculation used for Persistence, \gls{ARIMA} and \gls{CNN} detector (see Subsection \ref{sec:HTM}).
However, the comparatively poor detection performance also indicates a poor underlying forecast that is even outperformed by a trivial persistence forecast.
A potential reason could be insufficient adaption of the various model parameters to the dataset and scenario. 
Although the authors of \gls{HTM} claim the provided set of parameters to be the best for anomaly detection, it may not be sufficiently appropriate for the given scenario. However, as explained before, due to the low signal-to-noise ratio, it can be expected that even extensive parameter tuning will not result in a significantly better forecast compared to the persistence forecast.  

Fig. \ref{fig:F1_score_AUC_PR_line_plot} (c) shows the \textit{FAD} score for all detectors over the threshold $\tau$.
By comparing the $F_{1}$ score with the \textit{FAD} score, a shift between the optimal threshold $\tau_{\text{opt,F1}}$ and $\tau_{\text{opt,FAD}}$ towards smaller values can be noticed.
A smaller threshold increases the number of \glspl{TP} and results in earlier detection of an event.
At the same time, the number of \glspl{FP} increases as well.
However, as described in Subsection \ref{sec:metrics_detection} the \textit{FAD} score emphasises early event detection and weights \glspl{FP} low compared to \glspl{FN}, explaining the decrease of the optimal threshold. 
Based on the \textit{FAD} score an optimal threshold for the proposed Persistence detector of $\tau_{\text{opt,FAD}} = 0.16$ is determined. 
At $\tau_{\text{opt,FAD}}$ $191$ of $205$ \gls{FA} events (\SI{93}{\percent}) are detected by the Persistence detector, while $498$ data points of all $39827$ data points outside the event and rebound windows (\SI{1.25}{\percent}) are falsely declared and event. 

As previously described, for the dataset and scenario under investigation, more sophisticated models such as \gls{ARIMA} and \gls{CNN} only achieve minor improvements of the forecast compared to the persistence forecast. 
This also translates to a similar characteristic of the \textit{FAD} score curve over the threshold $\tau$. 
It can be inferred, that for the detection of fast-ramped \gls{FA} events an upper performance limit should exist at an \textit{FAD} score of roughly $\text{\textit{FAD}} \approx 0.7$. 
This performance limit can approximately be reached with the proposed Persistence detector (\textit{FAD}$_{\text{max}} = 0.69$). 
More advanced detection methods, such as the \gls{ARIMA} and \gls{CNN} detectors, only slightly improve the detection performance, but require a significantly higher maintenance and computational effort.
The  Persistence detector is therefore proposed to avoid frequent time and computation intensive model re-training. 
As described in Section \ref{sec:Requirements} this is considered a great advantage in a scenario of edge computing-based distributed event detection with time and resource constraints. 

\subsection{\gls{OS} classification of \gls{FA} events} \label{sec:results_classification}
In Fig. \ref{fig:Confusion_matrix} the confusion matrix for the \gls{EVM} model applied on the \gls{OS} test dataset is depicted.
Besides the two known event classes \gls{FA} and \gls{NO}, three unknown event classes are included in the test dataset, namely \gls{MP}, \gls{FV} and \gls{DU}, which are summarized as "unknown". 
The test dataset in total contains $63$ observations with $N_{\text{FA}}=21$, $N_{\text{NO}}=21$, $N_{\text{MP}}=7$, $N_{\text{FV}}=7$, $N_{\text{DU}}=7$. 
The test dataset has an openness of $O=\SI{24.7}{\percent}$. 
From Fig. \ref{fig:Confusion_matrix} it can be derived that the \gls{EVM} is able to correctly classify \SI{90}{\percent} of the \gls{FA} events and \SI{76}{\percent} of the \gls{NO} events.
Moreover, the \gls{EVM} successfully rejects \SI{71}{\percent} of all observations of the unknown classes. 
It can be concluded that the \gls{EVM} in principle is able to differentiate between \gls{FA} and \gls{NO} observations also in an \gls{OS} scenario with an acceptable performance.
However, the \gls{EVM} also erroneously classifies \SI{29}{\percent} of the observations from the unknown classes as \gls{FA} events, which reduces precision for \gls{FA} events.
Precision for \gls{NO} events is not affected by the unknown event classes.
Also the recall of the \gls{FA} and \gls{NO} event classes is negatively influenced, since \SI{10}{\percent} of the \gls{FA} events and \SI{19}{\percent} of the \gls{NO} events, respectively, are rejected.
%The matrix in numbers
%Horizontal target class
%Vertical output class
\def\myConfMat{{
{19,0,2},  %row 1
{1,16,4},  %row 2
{6,0,15},  %row 3
}}
\def\classNames{{"\smash{FA}","\smash{NO}","\smash{unknown}"}} %class names. Adapt at will
\def\numClasses{3} %number of classes. Could be automatic, but you can change it for tests.
\def\myScale{1.5} % 1.5 is a good scale. Values under 1 may need smaller fonts!
\begin{figure}[t]
\centering
\begin{tikzpicture}[
    scale = \myScale,
    %font={\scriptsize}, %for smaller scales, even \tiny may be useful
    ]

\tikzset{vertical label/.style={rotate=90,anchor=east}}   % usable styles for below
\tikzset{diagonal label/.style={rotate=45,anchor=north east}}

\foreach \y in {1,...,\numClasses} %loop vertical starting on top
{
    % Add class name on the left
    \pgfmathparse{-\y + {-0.1,-0.1,-0.05}[\y-1]}
    \node [anchor=east] at (0.4,\pgfmathresult) {\pgfmathparse{\classNames[\y-1]}\pgfmathresult}; 
    
    \foreach \x in {1,...,\numClasses}  %loop horizontal starting on left
    {
%---- Start of automatic calculation of totSamples for the column ------------   
    \def\totSamples{0}
    \foreach \ll in {1,...,\numClasses}
    {
        \pgfmathparse{\myConfMat[\y-1][\ll-1]}   %fetch next element
        \xdef\totSamples{\totSamples+\pgfmathresult} %accumulate it with previous sum
        %must use \xdef fro global effect otherwise lost in foreach loop!
    }
    \pgfmathparse{\totSamples} \xdef\totSamples{\pgfmathresult}  % put the final sum in variable
%---- End of automatic calculation of totSamples ----------------
    
    \begin{scope}[shift={(\x,-\y)}]
        \def\mVal{\myConfMat[\y-1][\x-1]} % The value at index y,x (-1 because of zero indexing)
        \pgfmathtruncatemacro{\r}{\mVal}   %
        \pgfmathtruncatemacro{\p}{round(\r/\totSamples*100)}
        \coordinate (C) at (0,0);
        \ifthenelse{\p<50}{\def\txtcol{black}}{\def\txtcol{white}} %decide text color for contrast
        \node[
            draw,                 %draw lines
            text=\txtcol,         %text color (automatic for better contrast)
            align=center,         %align text inside cells (also for wrapping)
            fill=bblue!\p,        %intensity of fill (can change base color)
            minimum size=\myScale*10mm,    %cell size to fit the scale and integer dimensions (in cm)
            inner sep=0,          %remove all inner gaps to save space in small scales
            ] (C) {\r\\\p\,\%};     %text to put in cell (adapt at will)
        %Now if last vertical class add its label at the bottom
        \ifthenelse{\y=\numClasses}{
        \node [] at ($(C)-(0,0.75)$) % can use vertical or diagonal label as option
        {\pgfmathparse{\classNames[\x-1]}\pgfmathresult};}{}
    \end{scope}
    }
}
%Now add x and y labels on suitable coordinates
\coordinate (yaxis) at (-0.3,0.25-\numClasses/2);  %must adapt if class labels are wider!
\coordinate (xaxis) at (0.5+\numClasses/2, -\numClasses-1.1); %id. for non horizontal labels!
\node [vertical label] at (yaxis) {True event class};
\node []               at (xaxis) {Predicted event class};
\end{tikzpicture}
    \caption{Confusion matrix of the \gls{EVM} model on the \gls{OS} classification test dataset with an openness $O$ = \SI{24.7}{\percent}.} \label{fig:Confusion_matrix}
\end{figure}
In general, Fig. \ref{fig:Confusion_matrix} demonstrates that the influence of the unknown classes on precision and recall of a class is higher compared to the influence of the other known class. 

\begin{figure}[b]
\centering
    \begin{tikzpicture}
\begin{axis}[
    width  = 1.03*\linewidth,
    height = 4.8cm,
    xlabel={Openness $O$ [-]},
    ylabel={$F_{1}$ score [-]},
    xticklabel={$\pgfmathprintnumber{\tick}\,\%$},
    major x tick style = transparent,
    major y tick style = transparent,
    legend pos=south west,
    ymajorgrids=true,
    xmajorgrids=true,
    legend cell align=left,
        legend style={
                at={(0.36,0.05)},
                anchor=south east,
                column sep=1ex}
]

\addplot[
    color=cornflowerblue,
    mark=*,
    line width=0.5mm,
    ]
    coordinates {(0,0.896) (10.56, 0.864) (18.35,0.864) (24.407,0.837)};

\addplot[
    color=sandybrown,
    mark=*,
    line width=0.5mm,
    ]
    coordinates {(0,0.905) (10.56, 0.839) (18.35,0.792) (24.407,0.742)};
    
    \legend{\gls{OS} EVM, \gls{CS} EVM}
    
\node[scale=0.78]  at (103,160) {+MP};
\node[scale=0.78]  at (179,160) {+MP,\,FV};
\node[scale=0.78]  at (236,160) {+MP,\,FV,\,DU};

\draw [loosely dashed,color=gray] (105.6, -8) -- (105.6,300);
\draw [loosely dashed,color=gray] (183.5, -8) -- (183.5,300);
\draw [loosely dashed,color=gray] (244.07, -8) -- (244.07,300);

\end{axis}
\end{tikzpicture}
    \caption{Comparison of the $F_{1}$ score between \gls{OS} and \gls{CS} \gls{EVM} model on the \gls{OS} classification problem with a varying openness $O$.} \label{fig:EVM_line_plot}
\end{figure}
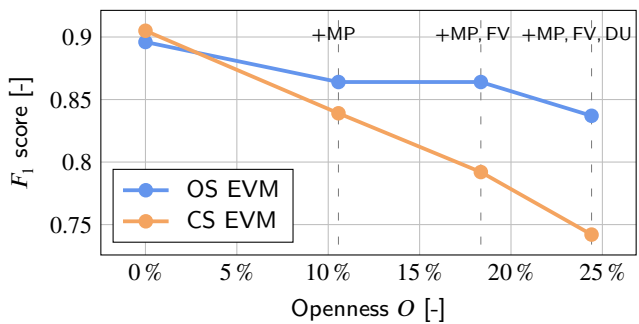

In order to investigate the benefit of applying an \gls{OS} classifier in the more realistic scenario with presence of unknown classes, performance is compared to a \gls{CS} classifier. 
For this purpose, the \gls{EVM} model is applied on the test dataset as both an \gls{OS} and \gls{CS} classifier.
In the \gls{CS} setting the rejection of observations with $\hat{P}(C_{l}|v')<0.9$ is deactivated and observations are classified according to $\hat{P}(C_{l}|v')$.
Moreover, in order to investigate the influence of the number of unknown classes, the comparison is conducted on a test dataset with increasing fractions of unknown classes.
In Fig. \ref{fig:EVM_line_plot} the comparison of the \gls{OS} and \gls{CS} \gls{EVM} for a varying openness $O$ of the test dataset is depicted. 
Performance is evaluated based on the $F_{1}$ score. 
For the \gls{CS} problem ($O=0$) the performance of the \gls{CS} classifier is slightly better compared to the \gls{OS} classifier, since the \gls{OS} classifier wrongly rejects some of the observations of the known classes. 
By adding \gls{MP} events as unknown class to the test dataset the openness of the test dataset increases to $O=\SI{10.56}{\percent}$. 
In this scenario the \gls{OS} classifier outperforms the \gls{CS} classifier.
While the \gls{OS} classifier is capable of rejecting observations from unknown classes, the \gls{CS} classifier assigns all observations of unknown classes to one of the known classes, resulting in a decreased precision.
However, the performance of the \gls{OS} \gls{EVM} decreases as well.
This is due to two reasons. 
First, not all observations from the unknown classes are successfully rejected. 
Second, being capable of rejecting observations can also lead to falsely rejected observations of known classes.
Nevertheless, for the investigated scenario of \gls{FA} event classification the rejection capability improves the performance compared to the \gls{CS} classifier already for the existence of only one unknown class. 
Extending the test dataset with the unknown \gls{FV} event class ($O=\SI{18.35}{\percent}$) has no influence on the performance of the \gls{OS} \gls{EVM}.
This can be explained by the specific characteristic of \gls{FV} events.
The number of zeros $n_{0}$ constitutes a strong differentiator for observations of the \gls{FV} event class, making it comparatively easy for the \gls{OS} \gls{EVM} to differentiate between \gls{FV} and the known \gls{FA} and \gls{NO} events.
Nevertheless, the $F_{1}$ score of the \gls{CS} classifier further decreases from ${F}_{1} = 0.839$ to ${F}_{1} = 0.792$ since all observations of the \gls{FV} event class are assigned to either the \gls{FA} or \gls{NO} event class.
In the final scenario ($O=\SI{24.7}{\percent}$) all unknown event classes are added to the test dataset, corresponding to the scenario described by Fig. \ref{fig:Confusion_matrix}. Adding the unknown \gls{DU} event class further decreases performance for both the \gls{OS} and \gls{CS} classifier.
However, while for the \gls{CS} \gls{EVM} the $F_{1}$ score decreases by $6.31$ percent, the \gls{OS} classifier only shows a decrease of $3.12$ percent.
In summary, the performance of the \gls{OS} \gls{EVM} decreased from ${F}_{1} = 0.896$ ($O=\SI{0}{\%}$) to ${F}_{1} = 0.837$ ($O=\SI{24.7}{\%}$), while the \gls{CS} classifier performance decreased from ${F}_{1} = 0.905$ ($O=\SI{0}{\%}$) to ${F}_{1} = 0.742$ ($O=\SI{24.7}{\%}$).
This demonstrates that the rejection capability of the \gls{OS} classifier allows maintaining the classification performance on a higher level, for increasing fractions of unknown classes. 
Nevertheless, also for the \gls{OS} classifier the performance deteriorates with additional unknown classes. 

To summarize, \gls{FA} events can be classified also in the more realistic \gls{OS} scenario and applying \gls{OS} classifiers can significantly improve performance under these conditions.
However, the classification performance of the \gls{EVM} is expandable, due to the very limited training data.
The size of the dataset constitutes a limitation of the presented study, since the small number of observations and classes prevent more comprehensive investigations.
Nevertheless, this study proves the fundamental feasibility of \gls{OS} classification of \gls{FA} events on real data. 
\section{Conclusion and future work} \label{sec:Conclusion}
This work demonstrates the fundamental feasibility of unsupervised detection and \gls{OS} classification of fast-ramped \gls{FA} events.
A data processing pipeline for \gls{FA} event identification is suggested, which combines both steps.
For unsupervised \gls{FA} event detection, a simple Persistence detector is proposed and implemented. The comparison with more complex and computationally expensive detection models demonstrates a similar performance and the existence of an upper performance limit.
Results indicate that the Persistence detector is particularly suitable for the specific class of \gls{FA} events.
As \gls{OS} classifier the \gls{EVM} is used. 
It is shown that the use of an \gls{OS} classifier significantly improves classification performance in the more realistic \gls{OS} scenario compared to a traditional \gls{CS} classifier.
Both the Persistence detector and \gls{EVM} classifier are selected with a view to an application in a distributed event detection architecture with time and resource constraints due to edge computing.
Their good performance demonstrates that main building blocks of the proposed pipeline can be realized with comparatively simple and lightweight methods that fulfill important requirements for an application in a distributed event detection architecture.

Given the fundamental proof of the main building blocks, a logical next step is the investigation of the coupling of the Persistence detector and \gls{EVM} classifier in the proposed \gls{EIP} for \gls{FA} events.
Moreover, for both the detection as well as classification step, several possible improvements could be investigated.
One direction could be the integration of additional regressors for unsupervised event detection such as temperature and solar radiation. 
For the \gls{OS} classification problem principle component analysis or other methods for dimensionality reduction could be applied to reduce the feature space dimension while retaining the majority of the information.
Finally, the proposed pipeline could be extended from \gls{FA} event identification to identification of multiple relevant events in active \glspl{DG}.
\section*{Acknowledgement}
This work is partly funded by the Innovation Fund Denmark (IFD) under File No. 91363 and by the INTERPRETER project,  which has received funding from the European Union’s Horizon 2020 research and innovation programme under grant agreement No. 864360.

\bibliographystyle{ieeetr}

\bibliography{References}

\section*{Appendix}\label{sec:appendix_B}

\begin{algorithm*}[b]
\caption{General procedure of the \gls{EIP} for \gls{FA} events.}  \label{algo:pipeline}
\begin{algorithmic}[1]
\For {new incoming data point $x_{\Delta,t}$}:

 \State \Call{persistence\_forecast}{$x_{\Delta,t}$} \Comment{Start of unsupervised event detection}
        \State \hspace{14pt} \Return $\hat{x}_{\Delta,t}$
    \If {$|\hat{x}_{\Delta,t}-x_{\Delta,t}|<\tau$}:
        \State Declare $x_{\Delta,t}$ normal behavior
        
    \Else:
        \State Declare $x_{\Delta,t}$ an event
        \State \Call{backward\_sampling}{$x_{\Delta,t}$} \Comment{Start of event sampling}
            \State  \hspace{14pt} \Return $\bm{x}_{\Delta,\text{bw}}=\{x_{\Delta,t-w_{\bm{x}}},...,x_{\Delta,t+e}\}$
        \State \Call{forward\_sampling}{$x_{\Delta,t}$} 
            \State \hspace{14pt} \textbf{if} another event at $x_{\Delta,t+a} \in \{x_{\Delta,t+1},...,x_{\Delta,t+w_{\bm{x}}}\}$ \textbf{then}:
            \State \hspace{28pt} \Return $\bm{x}_{\Delta,\text{fw}}=\{x_{\Delta,t-e},...,x_{\Delta,t+a+e}\}$
            \State \hspace{14pt} \textbf{else}:
            \State \hspace{28pt} \Return $\bm{x}_{\Delta,\text{fw}}=\{x_{\Delta,t-e},...,x_{\Delta,t+w_{\bm{x}}}\}$

        \For {$\bm{x}$ in [$\bm{x}_{\Delta,\text{bw}}$, $\bm{x}_{\Delta,\text{fw}}$]}:
            \State Calculate feature vector $v = [\mu_{\bm{x}}, \sigma_{\bm{x}},x_{\text{min}}, x_{\text{max}}, n_{0}, n_{\text{minmax}}]$
             \State \Call{extreme\_value\_machine}{$v$} \Comment{Start of \gls{OS} classification}
             \State \hspace{14pt} \Return $P(C_{flexibility}|v)$, $P(C_{normal\_behavior}|v)$
                \If {$P(C_{flexibility}|v) \geq P(C_{normal\_behavior}|v)$ \textbf{and} $\geq \rho$}:
                \State Declare sample $\bm{x}$ as flexibility activation event
                \ElsIf {$P(C_{flexibility}|v) \leq P(C_{normal\_behavior}|v)$ \textbf{and} $\geq \rho$}:
                \State Declare sample $\bm{x}$ as normal behavior
                \Else:
                \State Declare sample $\bm{x}$ as unknown event
                \EndIf
        \EndFor
    \EndIf
\EndFor

\end{algorithmic}
\end{algorithm*}

\end{document}